\documentclass[12pt]{article}
\usepackage{amsmath,amsthm,amsfonts,graphicx,epsfig}
\usepackage[usenames]{color}

\newcommand{\mbf}[1]{\mathbf #1}

\newcommand{\dbar}{\kern-.1em{\raise.8ex\hbox{ -}}\kern-.6em{d}}

\def\half{\mbox{$1\over2$}}

\def \be{\begin{equation}}
\def \ee{\end{equation}}
\begin{document}
\title{Swimming, Pumping and Gliding at low Reynolds numbers}
\author{O. Raz and J. E.  Avron
\\
Department of Physics\\ Technion, 32000 Haifa, Israel}

\date{\today}%
\maketitle 
\begin{abstract}

Simple, linear equations relate  microscopic swimmers to the
corresponding gliders and pumps. They have the following set of
consequences: The swimming velocity of free swimmers can be
inferred from the force on the tethered swimmer and vice versa; A
tethered swimmer dissipates more energy than a free swimmer; It is
possible to swim with arbitrarily high efficiency, but it is
impossible to pump with arbitrarily high efficiency  and finally
that pumping is geometric. We also solve several optimization
problems associated with swimming and pumping: The problem of
optimal anchoring for a certain class of swimmers that includes
the Purcell swimmer and the three linked spheres and the optimal
geometries of helices considered as swimmers and pumps.
\end{abstract}
\section{Introduction}
Low Reynolds numbers hydrodynamics governs the locomotion of tiny
natural swimmers such as micro-organisms and tiny artificial
swimmers, such as microbots.  It also governs micro-pumps
\cite{day-stone}, tethered swimmers \cite{Berg-pump,Cili_Pnas} and
gliding under the action of external forces. Our purpose here is
to discuss some interesting consequences of simple, yet general
and exact relations that relate pumping, swimming and gliding at
low Reynolds numbers. The equations are not new, and appear for
example in \cite{Yariv,Stone-Samuel}, where they serve to analyze
swimming at low Reynolds numbers. The interpretation of these
equations as a relation between three apparently different
objects: Swimmers, pumps and gliders, gives them a new and
different flavor and suggests new applications which we explore
below.

Our first tool is Eq.~(\ref{eq:skewed-linear-response}) which
relates the velocity of a free swimmer to the (torque and) force
needed to anchor it to a fixed location. This equation appears  in
the analysis of swimmers \cite{Yariv,Stone-Samuel}. It can be
applied to experiments on tethered swimmers
\cite{Pnas-swimming-efficiency,shoham,Hosoi} where it can be used
to infer the velocity of a free swimmers from measurements of the
force on tethered one (and vice versa) and when both are measured,
to an estimate of the gliding resistance matrix $M$ of
Eq.~(\ref{eq:M}).

Our second tool is Eq.~(\ref{eq:main-result}) which says that the
power needed to operate a pump --- a tethered swimmer --- is the
sum of the power invested by the corresponding free swimmer, and
the power needed to tow it. (The result holds for arbitrary
container of any shape or size). It is useful in studying and
comparing the efficiencies of swimmers and pumps. In particular,
it says that a tethered swimmer spends {\em more power} than a
free swimmer. As we shall see it also implies that it is
impossible to pump with arbitrary high efficiency.

Physically, both relations may be viewed as an expression of the
elementary observation that swimmers and pumps are the flip sides
of the same object: A tethered swimmer is a pump \cite{Berg-pump}
and an unbolted pump swims. This observation plays limited role in
high Reynolds numbers, because it does not lead to any useful
equations for the non-linear Navier Stokes equations. For
microscopic objects, where the equations reduce to the linear
Stokes equations, the observation translates to linear relations.

Formally, a pump is a swimmer with a distinguished point --- the
point of anchoring. The pump may have different properties
depending on the  point of anchoring, (we shall see an example
where it pumps in different directions depending on the anchoring
point). We shall solve a problem, posed by E. Yariv \cite{Yariv},
of how to find the optimal anchoring point for a certain class of
swimmers, which we call ``linear swimmers". The class includes the
Purcel swimmer \cite{Purcell}, the ``three linked spheres''
\cite{Golestanian} and pushmepullyou \cite{Avron-Oaknin-Kenneth}.

There are various notions of optimality that one needs to consider
in the study of swimmers and pumps. For swimmers, one notion of
optimality is to maximize the distance covered in one stroke. A
different notion is to minimize the dissipation in covering a
given distance at a given speed. Similarly, for a pump one may
want to maximize the momentum transfer to the fluid in one stroke,
or, alternatively, to minimize the dissipation for given total
momentum transfer and rate.  How are optimal swimmers related to
optimal pumps? We study this question for helices and show that
optimal swimmers and pumps have different geometry: There are four
optimal pitch angles depending on what one optimizes and whether
the helix is viewed as a swimmer or a pump.

\section{Anchoring as a choice of gauge}
Let us start by briefly discussing the gauge issues that arise
when considering swimming \cite{Wilczek-Shapere}, pumping and
gliding.

Fixing a gauge is already an issue for a glider. A glider is a
rigid body undergoing an Euclidean motion under the action of an
external force. There is no canonical way to decompose a general
Euclidean motion into a translation and a rotation
\cite{Landau-Lifshitz-m}. Such a decomposition requires choosing a
fiducial reference point in the body to fix the translation. In
the theory of rigid body a natural choice is the center of mass.
This is, however, {\em not} a natural choice at low Reynolds
number. This is because this is the limit when inertia plays no
role and the glider may be viewed as massless. Since the choice
becomes arbitrary, it may be viewed as a choice of gauge.

In the case of a swimmer one needs to fix a point arbitrarily and in
addition, to fix a fiducial frame. This is because a swimmer is a
deformable body and such a frame is required to fix the rotation
\cite{Wilczek-Shapere}. This is, again, a choice of gauge.

A pump is an anchored swimmer with a distinguished point and a
distinguished frame which are determined by the way the pump is
anchored. When we write equation that involve swimmers, pumps and
gliders, we pick the fiducial point and frame determine by the the
way the pump is anchored.

\section{Triality of swimmer pumps and gliders}

We shall first recall the derivation of the linear relation
between the 6 dimensional force-torque vector, ${\mathbf
F}_p=(F_p,N_p)$ which keeps a {\em pump} anchored with fixed
position and orientation, and the 6 dimensional
velocity--angular-velocity vector, ${\mathbf V}_s=(V_s,\omega_s)$
associated with the corresponding {\em autonomous swimmer}:
\begin{equation}\label{eq:skewed-linear-response}
    \mbf{F}_p= -M \,
      \mbf{V}_s,
\end{equation}
This holds for all times ($\mbf{F}_p, M$ and $\mbf{V}_s$ are time
dependent quantities). $M$ is a $6\times 6$  matrix of
linear-transport coefficients of the corresponding {\em glider}:
\begin{equation}\label{eq:linear-response}
      \mbf{F}_g
    = M\,  \mbf{V}_g
\end{equation}
namely, the corresponding rigid body, moving at (generalized)
velocity $\mbf{V}_g$ under the action of the (generalized) force
$\mbf{F}_g$. Note the change in sign. The matrix  $M$ depends on
the geometry of the body. It is a positive matrix of the form
\cite{Happel-Brenner}:
\begin{equation}\label{eq:M}
   M=       \left(\begin{array}{cc}
                 K & C \\
                 C^t & \Omega \\
               \end{array}
             \right)
\end{equation}
where $K, C,$ and $\Omega$ are $3\times 3$ real matrices. Note
that Eq.~(\ref{eq:linear-response}) fails in two dimensions
\cite{Landau-Lifshitz-fm}.



 Eq.~(\ref{eq:skewed-linear-response})  follows from the linearity
of the Stokes equations and the no-slip boundary conditions. Let
$\partial\Sigma$ denote the surface of the device.  Any  vector
field  on $\partial\Sigma$ can be decomposed into a deformation
and a rigid body motion as follows: Any rigid motion is of the
form $\mbf{v}_g={V}+\omega\times \mbf{x}$. Pick ${V}$ to be the
velocity of a fiducial point and $\omega$ the rotation of the
fiducial frame. The deformation field is then, by definition, what
remains when the rigid motion is subtracted from the given field
$\mbf{v}$.

Now, decompose $\mbf{v}_s$ the velocity field on the surface of a
swimmer, to a deformation and rigid-motion as above. The deformation
field can be identified with the velocity field at the surface of
the corresponding pump $\mbf{v}_p$ since the pump is anchored with
the fiducial frame that neither moves nor rotates. The remaining
rigid motion $\mbf{v}_g$ is then naturally identified with the
velocity field on the surface of the glider. The three vector fields
are then related by
\begin{equation}\label{eq:v-swimmer-pump-glider}
\mbf{v}_s=\mbf{v}_p+\mbf{v}_g, \quad \mbf{v}_g={V}_s+\omega_s\times
\mbf{x}
\end{equation}
where $V_s$ is (by definition) the swimming velocity and $\omega_s$
the velocity of rotation.

Each of the three velocity fields on $\partial\Sigma$, (plus the
no-slip zero boundary conditions on the surface of the container, if
there is one), uniquely determine the corresponding velocity field
and pressure $(v,p)$ throughout the fluid.
The stress tensor, $\pi_{ij}$, depends linearly on $(v,p)$
\cite{Landau-Lifshitz-fm} .

By the linearity of the Stokes, $\partial_j\pi_{ij}=0$, and
incompressibility $\partial_jv_j=0$ equations,  it is clear that
$v_s=v_g+v_p$ and $p_s=p_g+p_p$ and then also $\pi_s=\pi_p+\pi_g$.
Since $F_i=\int_{\partial\Sigma} \pi_{ij} dS_j$ is the drag force
acting on the device we get that the three force vectors are also
linearly related: $F_s=F_p+F_g$, and similarly for the torques. This
is summarized by the force-torque identity
$\mbf{F}_s=\mbf{F}_p+\mbf{F}_g$. Since the force and torque on an
autonomous Stokes swimmer vanish,
Eq.~(\ref{eq:skewed-linear-response}) follows from
Eq.~(\ref{eq:linear-response}).

Eq.~(\ref{eq:skewed-linear-response}) has the following
consequences:
\begin{itemize}
\item Micro-Pumping and Micro-Stirring is geometric: The momentum
and angular momentum transfer in a cycle of a pump, $\int{\mbf{F}_p
dt}$, is independent of its (time) parametrization. In particular,
it is independent of how fast the pump runs. This is because
swimming is geometric \cite{Purcell,Wilczek-Shapere} and the matrix
$M$ is a function of the pumping cycle, but not of its
parametrization.

\item
Scallop theorem  for pumps: One can not swim at low Reynolds numbers
with self-retracing strokes. This is known as the ``Scallop theorem"
\cite{Purcell}. An analog for pumps states that there is neither
momentum nor angular momentum transfer in a pumping cycle that is
self-retracing. This can be seen from the fact that $\mbf{V}_s \,
dt$ is balanced by $-\mbf{V}_s \, dt$ when the path is retraced, and
this remains true for $M\mbf{V}_s\, dt$.

\end{itemize}

We shall now derive an equation, originally due to
\cite{Stone-Samuel}, which relates the power expenditure of
swimmers, pumps and gliders. It follows from Lorentz reciprocity for
Stokes flows that says that if $(v_j, \pi_{jk})$ and $(v_j',
\pi_{jk}')$ are the velocity and stress fields for two solutions of
the Stokes equations in the domain $\Sigma$ then
\cite{Happel-Brenner}:
\begin{equation}\label{Lorenz reciprocity}
\int_{\partial \Sigma} v'_i\, \pi_{ij}\, dS_j=\int_{\partial \Sigma}
v_i\,\pi_{ij}'\, dS_j
\end{equation}

For the problem at hand, we may take $\partial \Sigma$ to be the
surface of our device (since the velocity fields vanish on the
rest of the boundary associated with the container). The area
element $dS$ is chosen normal to the surface and pointing into the
fluid. Now apply the Lorentz reciprocity to a pump and a swimmer
velocity fields and use Eq.~(\ref{eq:v-swimmer-pump-glider}) on
both sides. This gives
\begin{equation}\label{eq:lr2}
-P_s+\mbf{V}_s \cdot \mbf{F}_s  = -P_p-\mbf{V}_s \cdot \mbf{F}_p
\end{equation}
where $P_s$ is the  power invested by the swimmer and $P_p$ the
power invested by the pump. Since the force and torque on the
swimmer vanish, $\mbf{F}_s=0$,  we get, using
Eq.~(\ref{eq:skewed-linear-response}) a linear relation between the
powers:
\begin{equation}\label{eq:main-result}
P_p-P_s=-\mbf{V}_s \cdot  \mbf{F}_p=\mbf{V}_s \cdot M
\mbf{V}_s=P_g\ge 0
\end{equation}
$P_g$ is the power needed to tow the glider. Since both swimming and
towing require positive power, at any moment pumping is more costly
than swimming or dragging.

The linearity  of Eq.~(\ref{eq:main-result}) is a noteworthy, and
somewhat unexpected. Eq.~(\ref{eq:v-swimmer-pump-glider}) says
that the corresponding velocity fields are linearly related.
Since power at low Reynolds number is quadratic in the velocity a
linear relation between the powers is not what one may naively
expect.


The most interesting consequence of this relation which, at least
for us, was somewhat of a surprise, is that a pumps needs more
power than a swimmer and so the power consumed by a tethered
swimmer is actually an upper bound on the power consumed by it
when freed.

One of the remarkable facts about low Reynolds number swimming is
that even though the dynamics is governed by dissipation it is
still possible to swim with arbitrarily high efficiency
\cite{Avron-Oaknin-Kenneth,treadmilling}. For this to happen, the
swimming velocity should be non-zero, but the energy dissipation
of the swimmer, $P_s$, should be zero, thus $P_p=P_g\ge 0$. The
same cannot happened for pumps: for a pump to be with arbitrarily
hight efficiency, it must have non-zero momentum transfer to the
fluid and zero power. Since $M$ of Eq.~(\ref{eq:M}) is a strictly
positive matrix $P_g$ is quadratic in the force and can not vanish
if the pump transfers momentum to the fluid.


\begin{figure}
\begin{center}
\includegraphics[width=4cm]{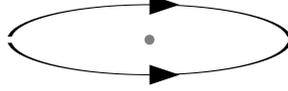}\\
\caption{A treadmiller transports its skin from head to back. The
treadmiller in the figure swims to the left while the motion of
its skin guarantees that the ambient fluid is left almost
undisturbed. If it is anchored at the gray point then it transfers
momentum to the fluid (to the right). The power of towing a frozen
treadmiller and of pumping almost coincide reflecting the fact
that a treadmiller  swims with little dissipation.
 }\label{fig:treadmiller}
\end{center}
\end{figure}

\section{Linear swimmers and optimal anchoring}

E. Yariv \cite {Yariv} posed the following problem: Find the
anchoring point which optimizes the momentum transfer to the
fluid. The general case is complicated. A class of swimmer for
which this question can be answered relatively easily is the class
of ``linear swimmers''  --- swimmers made of segments, so that the
velocity of different points on the same segment depend linearly
on the distance between the points. This class contains the three
linked spheres \cite{Golestanian}, the pushmepullyou
\cite{Avron-Oaknin-Kenneth}, Purcell's three linked swimmer
\cite{JFM-Stone,Tam-Hosoi}, the `N-Linked' swimmer
\cite{N-Linked-Rods}, but not, for example, the treadmiller
\cite{treadmilling}.

Shifting the anchoring point by $\vec{r}$, the resistance matrixes
$K$, $C$ and $\Omega$ of Eq.~\ref{eq:M} will be changed to
$K_r=K$, $C_r=C-KR$ and $\Omega_r=\Omega-RKR+CR-RC^t$
\cite{Happel-Brenner}, where $R$ is the antisymmetric matrix
associated  with the vector $\vec{r}$. The change in $V_s$ is
linear in $\vec{r}$, and $\omega_s$ is unchanged
\cite{Landau-Lifshitz-m}. Using
Eq.~(\ref{eq:skewed-linear-response}), the change in the force
$F_p$ (and hence in the linear-momentum transfer to the fluid) is
linear with $\vec{r}$, and the optimal point which maximized the
force must be at one of the edges of the segments.

(The change in the torque $T_p$ (and hence in the angular-momentum
transfer to the fluid) is quadratic in $\vec{r}$, so the maximum
can be either at the edges, or at the point in which the
differential of the angular momentum with respect to the anchoring
point is zero (in cases where there is such a point). In either
case, one has to check only a few points.)


A case in point is the ``three linked spheres" \cite{Golestanian}.
There are three candidates for the optimal anchoring point: The
three spheres. The two outer spheres are related by symmetry so the
two interesting cases are either anchoring on an outer sphere or in
the middle sphere. Detailed calculations show \cite{raz} that the
maximum momentum transfer will be when the anchoring point is on one
of the outer spheres. When the swimmer is anchored on the middle
sphere, the momentum transfer turns out to be {\em in the opposite
direction}. Calculation for the efficiency shows that the optimal
point in this case is also at the outer spheres.

\section{Helices as swimmers and pumps}
Eq.~(\ref{eq:skewed-linear-response}) and Eq.~(\ref{eq:main-result})
have an analogs for non-autonomous rigid swimmers such as a helix
rotating by the action of an external torque \cite{Nature}.
This is a case that is very easy to treat separately. From
Eq.~(\ref{eq:linear-response}) applied to the helix twice, once as
swimmer and once as a pump we get the analog of
Eq.~(\ref{eq:skewed-linear-response}):
\begin{equation}\label{eq:skewed-lin-res-helix}
    F_p=C\omega=-KV_s
\end{equation}
The analog of Eq.~(\ref{eq:main-result}) follows immediately from
the definition of the power $P=-\mbf{F}\cdot\mbf{V}$ and
Eq.~(\ref{eq:linear-response}) again:
\begin{equation}\label{eq:main-result-helix}
  P_p-P_s=\mbf{F}_p\cdot\mbf{V_p}-\mbf{F}_s\cdot\mbf{V}_s=-V_s\cdot
  F_p=P_g
\end{equation}
It follows from this that the difference in power between a swimmer
and a pump is minimized, for given swimming velocity, if the
swimming direction coincides with the smallest eigenvalue of $K$
which is the direction of optimal gliding.

The distinction between optimal pumps and optimal swimmers
\cite{Purcell_helix} can be nicely illustrated by the considering
the example of a rotating helix, or a screw. This motion has been
studied carefully in \cite{Lighthill,Higdon} whose analysis goes
beyond what we need here.  For a thin helix the slender-body
theory of Cox \cite{Cox} disposes of much of the hard work. Cox
theory has the small parameter $(\log \kappa)^{-1}$ where $\kappa$
is typically the ratio of the (large) radius of curvature of the
slender body, $r$ in the case of a helix, to its (small) diameter.
To leading order in $(\log \kappa)^{-1}$ the local force field on
the body is fixed by the {\em local} velocity field:
\begin{equation}\label{eq:cox}
dF(x)=k(\mbf{ t} ( \mbf{t}\cdot \mbf{v}) -2 \mbf{v}) dx,\quad
k=\frac{2\pi\mu}{\ln\kappa}
\end{equation}
$\mbf{ t}(x)$ is a unit tangent vector to the slender-body at $x$
and $\mbf{v}(x)$ the velocity of the point $x$ of the body. This
result may be interpreted as the statement that each line element
has twice the resistance in the transverse direction than the
resistance in the longitudinal direction, and that the motion of one
line element does not affect the force on another element (to
leading order).

Consider a helix  of radius $r$, pitch angle $\theta$ and total
length $\ell$. The helix is described by the parameterized curve
\begin{equation}\label{eq:helix}
    (r\cos\phi,r\sin\phi, t \sin\theta), \quad \phi=\frac{t}{r}\cos\theta,
    \quad t\in[0,\ell]
\end{equation}
Suppose the helix is being rotated at frequency $\omega$ about its
axis. Substituting the velocity field of a rotating helix, with an
unknown swimming velocity in the z-direction, into
Eq.~(\ref{eq:cox}), and setting the total force in the z-direction
to zero, fixes the swimming velocity. Dotting the force with the
velocity and integrating gives the power. This slightly tedious
calculation gives for the swimming velocity (along the axis) and
the power of swimming:
\begin{equation}\label{eq:h-swimmer}
\frac{V_{s}}{\omega r}=\frac{\sin2\theta}{3+\cos2\theta}, \quad
\frac{P_{s}}{k\ell \omega^{2}r^{2} }=\frac{4}{3+\cos 2\theta}
\end{equation}
Similarly, for the pumping force and power one finds
\begin{equation}\label{eq:h-pump}
\frac{F_{p}}{k\ell \omega r}=\sin\theta\cos\theta,\quad
\frac{P_{p}}{k\ell \omega^{2}r^{2}}=1+\sin^2\theta
\end{equation}
Eq.~(\ref{eq:h-swimmer}) and (\ref{eq:h-pump}) have the following
consequences for optimizing pumps and swimmers:
\begin{itemize} \item  Given $\omega r$, a swimmer velocity $V_s$  is
maximized at pitch angle $\theta=54.74^\circ$.

\item Given $\omega r$,  the pumping force $F_p$ is
maximized at $\theta= 45^\circ$.
\end{itemize}
Consider now optimizing {\em both}  the pitch angle $\theta$ and
rotation frequency $\omega$ so that the swimming velocity is
maximized for a given power. Namely
\begin{equation}\label{eq:optimal}
    \max_{\theta,\omega} \{V_s\ | \ P_s=const\}
\end{equation}
and similarly for pumping, except that $F_p$ replaces $V_s$  and
$P_p$ replaces $P_s$. A simple calculation shows that this is
equivalent to optimizing $V^2_s/P_s$ and $F^2_p/P_p$ with respect
to $\theta$. (These ratios are independent of $\omega$ and so
invariant under scaling time). One then finds:
\begin{itemize}
\item The  efficiency of swimming, $V_s^2/{P_s}$, is optimized at
$\theta= 49.9^\circ$. The efficiency  is proportional to
$(k\ell)^{-1}$ which favors small swimmers in less viscous media, as
one physically expects.

\item The efficiency of pumping, $F_p^2/{P_p}$, is optimized at
$\theta= 42.9^\circ$. The efficiency  is proportional to $(k\ell)$
which favors big pumps at more viscous media. Micro-pumps are
perforce inefficient.
\end{itemize}

There is a somewhat unrelated, yet insightful fact that one learns
from the above computation regarding the difference between motion
in a very viscous fluid and motion in a solid. The naive intuition
that the two are similar at very high viscosity would imply that a
helix moves like a cork-screw and so would move one pitch in one
turn. This is actually never the case, no matter how large $\mu$
is. In fact, the ratio of velocities of a helix to a cork-screw is
independent of $\mu$ and by Eq.~(\ref{eq:h-swimmer})
\begin{equation}\label{eq:cork}
\frac{V_s}{\omega r \sin\theta}
=\frac{\cos\theta}{1+\cos^2\theta}\le \half
\end{equation}
A helix needs at least two turns to advance the distance of its
threads.

\begin{figure}
\begin{center}
\includegraphics[width=6cm]{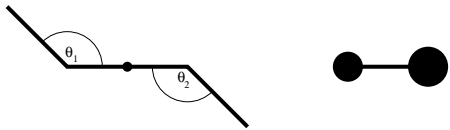}\\
\caption{A  Purcell three linked swimmer, left, controls the two
angles $\theta_1$ and $\theta_2$. When bolted it effectively
splits into two independent wind-shield wipers each of which is
self retracing. It will, therefore, not pump. Pushmepullyou,
right, controls the distance between the two spheres and the ratio
of their volumes. It can have arbitrarily large swimming
efficiency. }\label{fig:symetric_purcell}
\end{center}
\end{figure}

\section{Pumps that do no swim}

We have noted that  the ``three linked spheres'' can pump either
to the right or to the left depending on whether it is anchored on
the center sphere or on an external sphere.  There is, therefore,
an intermediate point where it will not pump.

There are also pumps that will not swim: Evidently, if the
swimming stroke is right-left and up-down symmetric, the swimmer
will not move by symmetry. It can, however, be bolted in a way
that breaks the symmetry to give an effective pump.  For example -
consider a right-left symmetric pushmepullyou
\cite{Avron-Oaknin-Kenneth}, where the  two spheres inflate and
deflate in phase. By symmetry, it will not swim, however,
anchoring in any point except the middle point will lead to net
momentum transfer.




\color{black}

\paragraph{ Acknowledgment} We thank A. Leshansky and O. Kenneth for
discussions, H.A Stone for useful correspondence and the  ISF and
the fund for promotion of research for support.


\end{document}